\documentclass[a4paper,11pt]{article}
\usepackage{jheppub}
\usepackage[T1]{fontenc}
\usepackage{graphicx}
\usepackage{array}
\usepackage{xcolor}
\usepackage{booktabs,multirow}
\usepackage[section]{placeins}
\usepackage{comment}
\usepackage{tikz}
\usepackage{hyperref}
\DeclareMathOperator{\Tr}{Tr}
\DeclareMathOperator{\tr}{tr}

\hypersetup{hidelinks,
	colorlinks=true,
	allcolors=blue!90!black,
	pdfstartview=Fit,
	breaklinks=true}

\newcommand{\avg}[1]{\langle \tr\, #1\rangle}

\newcommand{\ncsep}{0.34}
\newcommand{\ncletters}[1]{\foreach \l [count=\i] in {#1} {\node[font=\small] at ({\ncsep*(\i-1)},-0.22) {$\l$};}}
\newcommand{\ncblock}[4]{
	\foreach \i in {#1} {\draw[thick] ({\ncsep*(\i-1)},0) -- ({\ncsep*(\i-1)},#4);}
	\draw[thick] ({\ncsep*(#2-1)},#4) -- ({\ncsep*(#3-1)},#4);}
\newcommand{\ncbad}[4]{
	\foreach \i in {#1} {\draw[thick,dashed,red] ({\ncsep*(\i-1)},0) -- ({\ncsep*(\i-1)},#4);}
	\draw[thick,dashed,red] ({\ncsep*(#2-1)},#4) -- ({\ncsep*(#3-1)},#4);}

\title{Free cumulants and the analytic trajectory bootstrap for Yang-Mills matrix integrals}

\author{ Xinran Su}
\affiliation{Université Paris-Saclay, 91405 Orsay, France}
\emailAdd{xinransu.phys@gmail.com}

\abstract{
	We study the large $N$ limit of bosonic $D$-matrix Yang-Mills integrals
with the analytic trajectory bootstrap, which does not use positivity.
We improve the ansatz of the previous analytic trajectory bootstrap study~\cite{Li:2025ymi} by using the free cumulants, following a truncation proposed by Schwartz~\cite{Schwartz:1999cm}.
We take the free cumulants
as the unknowns,
and set the free cumulants of length greater than $K$ to zero.
The remaining unknowns are fixed by the loop equations.
We push the bootstrap to $K=12$ for $D=3,\dots,10$, using  the $\mathrm{O}(D)$ decomposition
and  the loop equations up to length $K+2$.
At $K=12$, the results for $\langle \tr XX\rangle$ deviate from Monte Carlo estimates by less than $0.2\%$ for $D\geq 6$.
For all the truncations $K=2,\dots,12$ and $D=3,\dots,10$, the corresponding single-matrix eigenvalue density is positive, although positivity is not imposed.
The truncation $K=4$ reproduces the leading and subleading orders of the $1/D$ expansion.
Imposing an ansatz for the large $D$ expansion of the free cumulants and solving the loop equations order by order in $1/D$, we derive the $1/D$ expansion of $\langle \tr XX\rangle$ up to order $D^{-4}$ relative to the leading term, two orders beyond Schwartz's result.
}

\begin{document}
\maketitle
\flushbottom

\section{Introduction}
\label{sec:intro}

Large $N$ matrix models provide nonperturbative definitions of string theory and M-theory,
in which spacetime is expected to emerge from the matrix degrees of freedom.
Two prominent examples are the BFSS (Banks-Fischler-Shenker-Susskind) matrix quantum mechanics,
proposed as a nonperturbative definition of M-theory~\cite{Banks:1996vh},
and the IKKT (Ishibashi-Kawai-Kitazawa-Tsuchiya) matrix model,
i.e., the ten-dimensional supersymmetric Yang-Mills matrix integral,
proposed as a nonperturbative definition of type IIB superstring theory~\cite{Ishibashi:1996xs}.
Their mass-deformed versions, such as the BMN (Berenstein-Maldacena-Nastase) model~\cite{Berenstein:2002jq} and the polarized IKKT model~\cite{Bonelli:2002mb,Hartnoll:2024csr,Martina:2025kwc},
have also been studied.

These supersymmetric matrix models are difficult to solve.
After integrating out the fermions, one obtains a Pfaffian, which is in general not real and positive.
This sign problem makes Monte Carlo simulations difficult,
and it also prevents a direct use of methods that rely on a positive measure.

The bosonic Yang-Mills matrix integrals are a simpler setting for the IKKT matrix model.
At large $N$, they are well defined without regularization for $D\geq 3$ matrices~\cite{Krauth:1998yu,Austing:2001bd,Austing:2001pk}.
In a previous work~\cite{Li:2025ymi}, we studied the bosonic $D$-matrix Yang-Mills integrals with the positivity bootstrap~\cite{Lin:2020mme,Kazakov:2021lel}.
It gives rigorous bounds,
and at the length cutoff $L_{\max}=12$ its precision is comparable to Monte Carlo simulations for some $D$.
However, the positivity constraints follow from the positivity of the measure.
For the supersymmetric matrix models, the measure is not always positive because of the Pfaffian,
so the positivity bootstrap cannot be applied directly.
This motivates a bootstrap method that does not use positivity.

\paragraph{The analytic trajectory bootstrap.}
In~\cite{Li:2025ymi}, this model was also studied with such a method, the analytic trajectory bootstrap~\cite{Li:2024ggr}.
There, the moments are described as analytic functions of the lengths of the words, called \textit{trajectories}, by an ansatz with finitely many parameters.
These analytic functions are closely related to resolvents and eigenvalue densities.
The parameters are then fixed by the loop equations, and the consistency conditions. The consistency conditions include: the cyclicity of the trace, the symmetries of the model,
the contraction limit (when one of the lengths goes to zero, a multi-length moment reduces to a moment with fewer lengths),
and the intersections of trajectories (two families of words that contain the same word must give the same value for it).
In~\cite{Li:2024ggr,Li:2025ymi}, the ansatz is motivated by the square-root behavior of the eigenvalue density near the endpoints of its support. However, the method is far less systematic than the positivity bootstrap, and the consistency conditions have to be imposed by hand.
In particular, the results at small $D$, such as $D=3$, are not well controlled~\cite{Li:2025ymi}.

\paragraph{The free cumulant truncation.}
A truncation of the loop equations based on free cumulants was proposed by Schwartz more than twenty years ago~\cite{Schwartz:1999cm}, based on the algebraic formulation of large $N$ matrix models by Halpern and Schwartz~\cite{Halpern:1999ia}.
The unknowns are the planar connected parts of the correlators,
which are the free cumulants in the language of free probability~\cite{Speicher:1994,Nica:2006}.
Schwartz assumed that the free cumulants decrease rapidly with their length,
and set all the free cumulants of length greater than a cutoff $K$ to zero.
For our model~\eqref{eq:action}, they obtained closed-form solutions at $K=2$ and $K=4$,
and the first two corrections of the large $D$ expansion at $K=6$.
They also observed a fast convergence with $K$ in matrix quantum mechanics.
The same truncation was recently used by Frenkel to evaluate variational ground states of matrix quantum mechanics~\cite{Frenkel:2026mqm}.

\paragraph{What is new in this work.}
In this work, we use the truncation of Schwartz as a new ansatz for the analytic trajectory bootstrap, and apply it to the Yang-Mills matrix integrals.
Here the ansatz is made for the free cumulants: all the free cumulants of length greater than $K$ vanish. For a single matrix, this determines the resolvent through the R-transform, so it can also be viewed as an ansatz for the trajectories (section~\ref{sec:trajectory}).
Compared with the ansatz of~\cite{Li:2025ymi}, this ansatz has two advantages.
First, the consistency conditions above are satisfied automatically, because all the moments are computed from the same free cumulants by the moment-cumulant formula.
Second, it can be improved systematically by increasing $K$.  For more detailed discussion, see section~\ref{sec:consistency}.
The lowest truncation $K=2$ coincides with the leading ansatz of~\cite{Li:2025ymi}.
We obtain the following results.

\begin{enumerate}
	\item \emph{$\mathrm{O}(D)$ parametrization and higher truncations.}
	We parametrize the free cumulants by $\mathrm{O}(D)$-invariant tensors.
	We use all the loop equations up to length $K+2$ and solve the overdetermined system by least squares.
	In this way, we push the truncation to $K=12$ for $D=3,\dots,10$.
	\item \emph{Comparison with rigorous bounds and Monte Carlo.}
	At $K=12$, $\avg{XX}$ deviates from the Monte Carlo estimates by $9\%$, $1.5\%$ and $0.4\%$ for $D=3,4,5$,
	and by less than $0.2\%$ for $D\geq 6$.
	For every $D$, the results lie inside the projections of the $L_{\max}=12$ positivity islands of~\cite{Li:2025ymi}.
	\item \emph{Eigenvalue density.}
	For finite $K$, we can derive the single-matrix resolvent.
	Although positivity is not imposed, the corresponding eigenvalue density is positive in $D=3,\dots,10$ and $K=2,\dots,12$. 
	
	\item \emph{Large $D$ expansion.}
	The truncation $K=4$ reproduces all the leading and subleading results of the $1/D$ expansion in~\cite{Li:2025ymi}.
		Beyond $K=4$, we obtain the $1/D$ expansion in two ways.
	Numerically, we find the least-squares solution of the truncated loop equations at several large values of $D$ and interpolate in $1/D$.
	This gives the coefficients of the $1/D$ corrections up to $K=10$.
	Analytically, we impose an ansatz for the large $D$ expansion of the free cumulants, keep the free cumulants of length $\leq K'$, and solve the loop equations order by order in $1/D$.
	This gives exact rational coefficients up to $K'=12$.
	The two methods give the same coefficients.
	The coefficient of $D^{-2}$ in $\sqrt{2D}\,\avg{XX}$ reproduces Schwartz's $n=6$ result,
	and the coefficients of $D^{-3}$ and $D^{-4}$ are new.
\end{enumerate}

The paper is organized as follows.
In section~\ref{sec:setup}, we introduce the model, the loop equations, the $\mathrm{O}(D)$ decomposition and the benchmark data.
In section~\ref{sec:method}, we formulate the free cumulant ansatz for the analytic trajectory bootstrap.
We first summarize the procedure, and then explain each ingredient.
In section~\ref{sec:results}, we present the results:
the closed-form solutions
of the lowest truncations, the numerical results up to $K=12$, the eigenvalue densities and the analytic large $D$ expansion.
In section~\ref{sec:discussion}, we discuss the limitations and some future directions.
Appendix~\ref{app:cost} lists the sizes of the systems,
and appendix~\ref{app:redundancy} discusses a redundancy of the cumulant parameters at small $D$.

\section{Preliminaries}

\label{sec:setup}

In this section, we briefly review the model, the loop equations, the $\mathrm{O}(D)$ decomposition of the moments,
and the results that we use as benchmarks. For more details, see~\cite{Li:2025ymi}.
\subsection{Model and notation}

The bosonic Yang-Mills matrix integral is defined by the action
\begin{equation}\label{eq:action}
	S=-\frac{N}{4}\Tr\,\big([X^\mu,X^\nu][X^\mu,X^\nu]\big) ,
\end{equation}
where $X^1,\dots,X^D$ are traceless Hermitian $N\times N$ matrices.
We use the conventions of~\cite{Li:2025ymi}.
The normalized trace is $\tr=\Tr/N$, and the expectation values are normalized such that $\langle \tr\,\mathbb{I}\rangle=1$.
We only consider the large $N$ limit.

The observables are the expectation values of single-trace words,
\begin{equation}\label{eq:expectation}
	\big\langle \tr\, X^{\mu_1}\cdots X^{\mu_L}\big\rangle
	\equiv
	\frac{\int \Big(\prod_{\mu=1}^{D} dX^\mu\Big)\,
		\tr\big(X^{\mu_1}\cdots X^{\mu_L}\big)\,e^{-S}}{	\int \Big(\prod_{\mu=1}^{D} dX^\mu\Big)\,e^{-S}},
\end{equation}
where $L$ is the length of the word.
We use the letters $X,Y,Z,W$ to denote distinct matrices,
for instance $\avg{XXYY}\equiv\langle\tr\, X^1X^1X^2X^2\rangle$.
The action has the following symmetries:
\begin{itemize}
	\item the $\mathrm{O}(D)$ transformations of $X^\mu$, which include the relabelling of the matrices
	and the reflection $X^\mu\to -X^\mu$ of a single matrix,
	so a word vanishes if a letter appears an odd number of times;
	\item the transposition $X^\mu \to (X^\mu)^T$, which reverses the order of a word.
\end{itemize}
Together with the cyclicity of the trace, these symmetries identify many words.
We assume that they are not spontaneously broken.

\subsection{Loop equations}

The loop equations relate words of different lengths.
They follow from integrating a total derivative in \eqref{eq:expectation}.
Fix a word $\mathcal{O}\equiv X^{\mu_1}\cdots X^{\mu_L}$ and an index $\nu$.
Assuming that the boundary terms vanish, we have
\begin{equation}\label{eq:total-derivative}
	\int \Big(\prod_{\mu=1}^{D} dX^\mu\Big)\,
	\frac{\partial}{\partial (X^{\nu})_{ij}}\big(\mathcal{O}_{ij}\, e^{-S}\big)=0 ,
\end{equation}
with $i,j$ summed over $1,\dots,N$.
At large $N$, the expectation value of a product of two traces factorizes,
and \eqref{eq:total-derivative} becomes
\begin{equation}\label{eq:loop}
	\big\langle \tr\,\mathcal{O}\big(X^{\nu}X^{\rho}X^{\rho}+X^{\rho}X^{\rho}X^{\nu}-2\,X^{\rho}X^{\nu}X^{\rho}\big)\big\rangle
	=\sum_{p=1}^{L}\delta^{\mu_p\nu}\,
	\big\langle \tr\,\mathcal{O}^{(p)}_{\mathrm{l}}\big\rangle\,
	\big\langle \tr\,\mathcal{O}^{(p)}_{\mathrm{r}}\big\rangle .
\end{equation}
Here $\mathcal{O}^{(p)}_{\mathrm{l}}\equiv X^{\mu_1}\cdots X^{\mu_{p-1}}$ and $\mathcal{O}^{(p)}_{\mathrm{r}}\equiv X^{\mu_{p+1}}\cdots X^{\mu_L}$.
The simplest example is $\mathcal{O}=X$ and $X^\nu=X$:
\begin{equation}\label{eq:loop-X}
	\avg{XXYY}-\avg{XYXY}=\frac{1}{2(D-1)} .
\end{equation}
We say that a loop equation has length $L+3$, the maximal length of the words in it.

\subsection{\texorpdfstring{$\mathrm{O}(D)$}{O(D)} decomposition}
\label{sec:ODdecomposition}
In this subsection, we review the $\mathrm{O}(D)$ decomposition of the moments in~\cite{Li:2025ymi}.
A moment with general indices is an $\mathrm{O}(D)$-invariant tensor.
Therefore, it can be written as a linear combination of products of Kronecker deltas,
\begin{equation}\label{eq:A-decomp}
	\big\langle \tr\, X^{\mu_1}X^{\mu_2}\cdots X^{\mu_L}\big\rangle=\sum_{M}A_{M}\prod_{(s,t)\in M}\delta^{\mu_s\mu_t},
\end{equation}
where $M$ runs over the perfect matchings of the $L$ positions, i.e., the ways to divide them into $L/2$ pairs $(s,t)$.
As in~\cite{Li:2025ymi}, we label a coefficient $A_M$ by the pairs of $M$,
e.g., $A_{1,2,3,4}$ is the coefficient of $\delta^{\mu_1\mu_2}\delta^{\mu_3\mu_4}$.
The simplest examples are
\begin{align}
	\big\langle \tr\, X^{\mu_1}X^{\mu_2}\big\rangle&= A_{1,2}\,\delta^{\mu_1\mu_2},\\
	\big\langle \tr\, X^{\mu_1}X^{\mu_2}X^{\mu_3}X^{\mu_4}\big\rangle
	&= A_{1,2,3,4}\,\delta^{\mu_1\mu_2}\delta^{\mu_3\mu_4}+A_{1,4,2,3}\,\delta^{\mu_1\mu_4}\delta^{\mu_2\mu_3}
	+ A_{1,3,2,4}\,\delta^{\mu_1\mu_3}\delta^{\mu_2\mu_4}.\label{eq:A-decomp-4}
\end{align}
For concrete words, \eqref{eq:A-decomp} keeps only the matchings that pair equal letters.
For example,~\eqref{eq:A-decomp-4} gives
\begin{equation}\label{eq:A-words}
	\avg{XXYY}=A_{1,2,3,4},\qquad \avg{YXXY}=A_{1,4,2,3},\qquad \avg{XYXY}=A_{1,3,2,4}.
\end{equation}
The coefficients $A_M$ are not all independent.
By the cyclicity of the trace, $\avg{YXXY}=\avg{XXYY}$, so~\eqref{eq:A-words} implies $A_{1,4,2,3}=A_{1,2,3,4}$.
In general, the cyclicity of the trace and the reversal symmetry imply that
two matchings related by a rotation or a reflection of the positions on a circle have the same coefficient.
We denote the class of $M$ under these operations by $[M]$.
The independent coefficients are $A_{[M]}$, one for each class,
and we label them by the pairs of one matching in the class.
For instance,~\eqref{eq:A-decomp-4} becomes
\begin{equation}
	\big\langle \tr\, X^{\mu_1}X^{\mu_2}X^{\mu_3}X^{\mu_4}\big\rangle
	= A_{1,2,3,4}\,\big(\delta^{\mu_1\mu_2}\delta^{\mu_3\mu_4}+\delta^{\mu_1\mu_4}\delta^{\mu_2\mu_3}\big)
	+ A_{1,3,2,4}\,\delta^{\mu_1\mu_3}\delta^{\mu_2\mu_4},
\end{equation}
and $\avg{XXXX}=2A_{1,2,3,4}+A_{1,3,2,4}$.
At small $D$, the products of deltas in~\eqref{eq:A-decomp} are not linearly independent.
In general, the products of $L/2$ deltas are linearly dependent if $L/2>D$,
so the coefficients $A$ of length $L>2D$ are not uniquely determined by the moments.
This redundancy is discussed in appendix B of~\cite{Li:2025ymi}, and its analogue for the free cumulants in appendix~\ref{app:redundancy}.

\subsection{Benchmarks}

\label{sec:benchmarks}

We compare the results of this work with three types of data:
the large $D$ expansion, the positivity bootstrap and Monte Carlo simulations.

\subsubsection{\texorpdfstring{Large $D$}{Large D} expansion}
As derived in \cite{Hotta:1998en}, the model admits a systematic $1/D$ expansion.
The results up to the subleading order were obtained in~\cite{Hotta:1998en,Li:2025ymi}. For example,
\begin{equation}\label{eq:largeD-sub}
	\avg{XX}=\frac{1}{\sqrt{2}}D^{-1/2}+\frac{7}{6\sqrt{2}}D^{-3/2}+O(D^{-5/2}),\qquad
	\avg{XXXX}=\frac{1}{D}+\frac{5}{2D^2}+O(D^{-3}).
\end{equation}
The results for many other words are listed in~\cite{Li:2025ymi}.

\subsubsection{Positivity bootstrap and Monte Carlo}
We compare our results with two sets of data.
The positivity bootstrap from~\cite{Li:2025ymi} at $L_{\max}=12$ gives rigorous intervals for $\avg{XX}$ and $\avg{XXXX}$,
and Monte Carlo (MC) simulations give estimates of these two moments and of some mixed words.
We use and adapt the code from~\cite{Jha:2021exo} to run MC simulations.
Table~\ref{tab:XX} reproduces the positivity intervals and the MC estimates of $\avg{XX}$ and $\avg{XXXX}$.

\section{Analytic trajectory bootstrap}
\label{sec:method}

In this section, we formulate the analytic trajectory bootstrap with the free cumulant ansatz.
The procedure consists of four steps:
\begin{enumerate}
	\item \emph{Change of variables.} Replace the moments by the free cumulants,
	which are related to the moments by the noncrossing moment-cumulant formula.
	This step is exact (section~\ref{sec:cumulant-def}).
	\item \emph{Symmetry reduction.} Use the $\mathrm{O}(D)$ symmetry to write each free cumulant of length $L$
	as a linear combination of products of Kronecker deltas. The coefficients $\theta_{[M]}$ are the unknown parameters (section~\ref{sec:ODforkappa}).
	\item \emph{Truncation.} Set all the free cumulants of length greater than $K$ to zero.
	This is the only approximation used in the bootstrap.
	Then every moment is a polynomial in the finitely many parameters $\theta_M$ of length $L\leq K$ (section~\ref{sec:truncation}).
	
	\item \emph{Loop equations.} Substitute these polynomials into the loop equations up to length $K+2$,
	and solve for the parameters.
	For $K\leq 4$, the system can be solved in closed form (section~\ref{sec:low}).
	For $K\geq 6$, there are more equations than unknowns, and we solve the system by least squares (section~\ref{sec:high}).
\end{enumerate}
The idea behind the truncation is that the free cumulants measure the interaction among the matrices,
and longer free cumulants are suppressed at large $D$. We will also show that this truncation can be interpreted as an ansatz for the trajectories (section~\ref{sec:truncation}).
Below we explain the four steps one by one.
The results are presented in section~\ref{sec:results}.
\subsection{Free cumulants and noncrossing partitions}
\label{sec:cumulant-def}

The free cumulants are the analogues of the connected correlators in the large $N$ limit.
In ordinary probability theory, a moment is the sum of the products of connected correlators (cumulants)
over all the ways to split the variables into groups.
At large $N$, only planar diagrams contribute.
Accordingly, a single-trace moment is the sum over the ways to split the letters of the word into groups
such that the groups do not cross each other.
The connected part of the whole word is the free cumulant.
Below we make this statement precise and give explicit examples.

\begin{figure}[htbp]
	\centering
	\scriptsize
	\begin{tabular}{l@{\hspace{1mm}}c@{\hspace{1mm}}c@{\hspace{1mm}}c@{\hspace{1mm}}c@{\hspace{1mm}}c}
		$\avg{XXYY}$ &
		\begin{tikzpicture}[baseline=0]\ncletters{X,X,Y,Y}\ncblock{1,2}{1}{2}{0.52}\ncblock{3,4}{3}{4}{0.52}\end{tikzpicture} &
		\begin{tikzpicture}[baseline=0]\ncletters{X,X,Y,Y}\ncblock{1,2,3,4}{1}{4}{0.68}\end{tikzpicture} & & & \\[1mm]
		& $\kappa(XX)\kappa(YY)$ & $\kappa(XXYY)$ & & & \\
		& $=a^2$ & $=b$ & & & \\[4mm]
		$\avg{XYXY}$ &
		\begin{tikzpicture}[baseline=0]\ncletters{X,Y,X,Y}\ncbad{1,3}{1}{3}{0.83}\ncbad{2,4}{2}{4}{0.52}\end{tikzpicture} &
		\begin{tikzpicture}[baseline=0]\ncletters{X,Y,X,Y}\ncblock{1,2,3,4}{1}{4}{0.68}\end{tikzpicture} & & & \\[1mm]
		& crossing & $\kappa(XYXY)$ & & & \\
		& (excluded) & $=c$ & & & \\[4mm]
		$\avg{XXYYZZ}$ &
		\begin{tikzpicture}[baseline=0]\ncletters{X,X,Y,Y,Z,Z}\ncblock{1,2}{1}{2}{0.52}\ncblock{3,4}{3}{4}{0.52}\ncblock{5,6}{5}{6}{0.52}\end{tikzpicture} &
		\begin{tikzpicture}[baseline=0]\ncletters{X,X,Y,Y,Z,Z}\ncblock{1,2}{1}{2}{0.52}\ncblock{3,4,5,6}{3}{6}{0.52}\end{tikzpicture} &
		\begin{tikzpicture}[baseline=0]\ncletters{X,X,Y,Y,Z,Z}\ncblock{1,2,3,4}{1}{4}{0.52}\ncblock{5,6}{5}{6}{0.52}\end{tikzpicture} &
		\begin{tikzpicture}[baseline=0]\ncletters{X,X,Y,Y,Z,Z}\ncblock{1,2,5,6}{1}{6}{0.83}\ncblock{3,4}{3}{4}{0.45}\end{tikzpicture} &
		\begin{tikzpicture}[baseline=0]\ncletters{X,X,Y,Y,Z,Z}\ncblock{1,2,3,4,5,6}{1}{6}{0.68}\end{tikzpicture} \\[1mm]
		& $\kappa(XX)\kappa(YY)\kappa(ZZ)$ & $\kappa(XX)\kappa(YYZZ)$ & $\kappa(XXYY)\kappa(ZZ)$ & $\kappa(XXZZ)\kappa(YY)$ & $\kappa(XXYYZZ)$ \\
		& $=a^3$ & $=ab$ & $=ab$ & $=ab$ & $=u_1$
	\end{tabular}
	\caption{Partitions in the moment-cumulant formula~\eqref{eq:mom-cum} for the words $XXYY$, $XYXY$ and $XXYYZZ$.
		Each word is written on a line, and the letters in the same block are connected by lines above the word.
		The product of free cumulants of the blocks is written below each diagram.
		All the contributing nonzero noncrossing partitions are shown.
		The red dashed diagram is intended to show a crossing partition, which does not appear in~\eqref{eq:mom-cum}.
		Summing each row gives~\eqref{eq:ex-XXYY}, \eqref{eq:ex-XYXY} and~\eqref{eq:ex-XXYYZZ}.}
	\label{fig:partitions}
\end{figure}
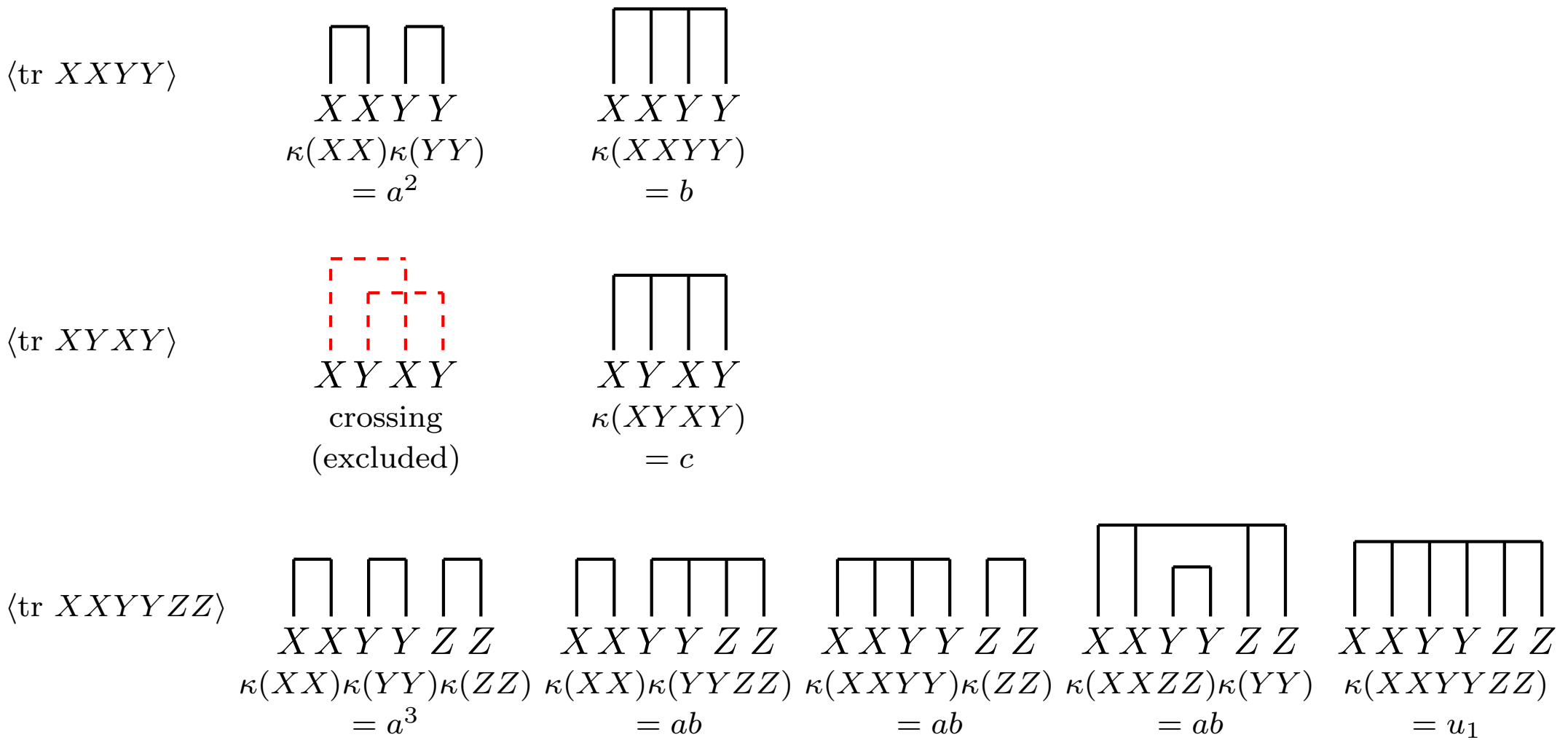

\paragraph{Partitions of a word.}
Consider a word $X^{\mu_1}X^{\mu_2}\cdots X^{\mu_L}$ inside a trace.
A partition of the word divides its $L$ letters into non-empty groups, called blocks.
Each block defines a shorter word, formed by the letters in the block in their original order.
A partition can have any number of blocks between $1$ and $L$.
For example, the word $XXYY$ can be kept as a single block $XXYY$.
It can be divided into the two blocks $XX$ (the first two letters) and $YY$ (the last two letters).
It can also be divided into two blocks that both read $XY$,
by grouping the first letter either with the fourth letter or with the third one.
However, because of the $\mathrm{O}(D)$ symmetry, in particular the reflection $X^\mu\to -X^\mu$ of a single matrix,
a block contributes to the moment-cumulant formula below only if each letter appears an even number of times in it.
Therefore, the two partitions into $XY$ and $XY$ do not contribute.
Below we only consider partitions whose blocks satisfy this condition.
We draw a partition by writing the word on a line and connecting the letters in the same block by lines above the word,
as in figure~\ref{fig:partitions}.

\paragraph{Noncrossing partitions.}
A partition is called noncrossing if these lines can be drawn without crossing each other.
For example, in the word $XYXY$, the partition into the block of the two $X$'s and the block of the two $Y$'s is crossing,
as shown by the red dashed lines in figure~\ref{fig:partitions}.
Among the partitions of $XXYY$ mentioned above, the one that groups the first letter with the third one is crossing, and the others are noncrossing.

\paragraph{Moment-cumulant formula.}
We label the positions in the word by $1,\dots,L$.
A partition $\pi$ is then a set of disjoint non-empty blocks whose union is $\{1,\dots,L\}$,
and we write each block as an increasing list of positions $\{p_1<p_2<\dots<p_r\}$.
In this notation, the partition is noncrossing if there are no positions $p<p'<p''<p'''$ such that $p,p''$ belong to one block and $p',p'''$ belong to another block.
We denote the set of noncrossing partitions of $L$ positions by $\mathrm{NC}(L)$.
The free cumulants $\kappa$ are defined by the moment-cumulant formula~\cite{Speicher:1994,Nica:2006}
\begin{equation}\label{eq:mom-cum}
	\big\langle \tr\, X^{\mu_1}X^{\mu_2}\cdots X^{\mu_L}\big\rangle=\sum_{\pi\in \mathrm{NC}(L)}\ \prod_{\{p_1<\dots<p_r\}\in\pi}\kappa\big(X^{\mu_{p_1}}X^{\mu_{p_2}}\cdots X^{\mu_{p_r}}\big).
\end{equation}
In words, each noncrossing partition contributes the product of the free cumulants of the shorter words formed by its blocks,
with the letters of each block kept in their original order.

The partition with a single block contributes $\kappa(X^{\mu_1}\cdots X^{\mu_L})$ itself,
and all the other partitions contain only shorter words.
Therefore,~\eqref{eq:mom-cum} can be solved for the free cumulant of the whole word,
\begin{equation}\label{eq:cum-recursion}
	\kappa(X^{\mu_1}\cdots X^{\mu_L})=\big\langle \tr\, X^{\mu_1}\cdots X^{\mu_L}\big\rangle
	-\sum_{\substack{\pi\in \mathrm{NC}(L)\\ \pi\neq\{\{1,\dots,L\}\}}}\ \prod_{\{p_1<\dots<p_r\}\in\pi}\kappa\big(X^{\mu_{p_1}}\cdots X^{\mu_{p_r}}\big).
\end{equation}
Starting from the shortest words, \eqref{eq:cum-recursion} determines all the free cumulants from the moments,
and~\eqref{eq:mom-cum} determines all the moments from the free cumulants.
In the language of planar diagrams, the free cumulants are the planar connected parts~\cite{Cvitanovic:1981yn}.
They are the unknowns used by Halpern and Schwartz~\cite{Halpern:1999ia,Schwartz:1999cm}.

\paragraph{Examples.}
We introduce the notation
\begin{equation}\label{eq:abc}
	a\equiv\kappa(XX),\qquad b\equiv \kappa(XXYY),\qquad c\equiv\kappa(XYXY),
\end{equation}
where $X$ and $Y$ are distinct matrices.
Figure~\ref{fig:partitions} shows the contributing partitions of a few words.
In terms of free cumulants, they read
\begin{align}
	\avg{XX}&=\kappa(XX)=a, \label{eq:ex-XX}\\
	\avg{XXYY}&=\kappa(XX)\,\kappa(YY)+\kappa(XXYY)=a^2+b, \label{eq:ex-XXYY}\\
	\avg{XYXY}&=\kappa(XYXY)=c, \label{eq:ex-XYXY}\\
	\avg{XXYYZZ}&=\kappa(XX)\,\kappa(YY)\,\kappa(ZZ)+\kappa(XX)\,\kappa(YYZZ)+\kappa(XXYY)\,\kappa(ZZ)\nonumber\\
	&\quad+\kappa(XXZZ)\,\kappa(YY)+\kappa(XXYYZZ)=a^3+3ab+u_1, \label{eq:ex-XXYYZZ}
\end{align}
where $u_1\equiv\kappa(XXYYZZ)$ will be discussed in section~\ref{sec:ODforkappa}.
Conversely,~\eqref{eq:cum-recursion} gives, e.g., $b=\avg{XXYY}-\avg{XX}^2$ and $c=\avg{XYXY}$.

\paragraph{Relation to freeness.}
The meaning of the free cumulants becomes clear in the free limit.
At the leading order of the large $D$ expansion~\cite{Hotta:1998en}, the matrices are free in the sense of Voiculescu~\cite{Voiculescu:1992}, i.e.,
\begin{equation}\label{eq:freeness}
	\Big\langle \tr
	\big[(X^{\mu_1})^{i_1} - \big\langle \tr\, (X^{\mu_1})^{i_1} \big\rangle\big]
	\cdots
	\big[(X^{\mu_m})^{i_m} - \big\langle \tr\, (X^{\mu_m})^{i_m} \big\rangle\big]
	\Big\rangle = 0,
\end{equation}
where neighboring matrices are distinct, $X^{\mu_j}\neq X^{\mu_{j+1}}$ for $j=1,\dots,m-1$ and $X^{\mu_m}\neq X^{\mu_1}$.\footnote{Freeness is the analogue of statistical independence for noncommuting variables~\cite{Voiculescu:1992,Nica:2006}.
	It concerns only the relations among different matrices, and it does not require each matrix to be Gaussian.
}
Applied recursively, \eqref{eq:freeness} expresses every mixed moment in terms of single-matrix moments.
For example, with $m=2$ it gives $\avg{XXYY}=\avg{XX}\,\avg{YY}$,
and with $m=4$ and all powers equal to one it gives $\avg{XYXY}=0$.
Comparing with~\eqref{eq:ex-XXYY} and~\eqref{eq:ex-XYXY}, we see that these relations are equivalent to $b=0$ and $c=0$.

In general, freeness is equivalent to the vanishing of all mixed free cumulants,
i.e., the free cumulants of words that contain at least two distinct matrices~\cite{Nica:2006,Speicher:1994}.
Therefore, the mixed free cumulants measure the deviation from freeness, in other words, the effects of the interaction among the matrices.
The single-matrix free cumulants $\kappa(X\cdots X)$ are not constrained by freeness.
At the large $D$ saddle point, each matrix is Gaussian, so its single-matrix free cumulants of length $\geq 4$ also vanish~\cite{Nica:2006}.
This is the starting point of the truncation in section~\ref{sec:truncation}:
we keep the free cumulants up to a finite length, and neglect the longer ones.

\subsection{\texorpdfstring{$\mathrm{O}(D)$}{O(D)} parametrization for free cumulants}
\label{sec:ODforkappa}
We now apply the $\mathrm{O}(D)$ decomposition of section~\ref{sec:ODdecomposition} to the free cumulants.
A free cumulant with general indices, $\kappa(X^{\mu_1}\cdots X^{\mu_L})$, is an $\mathrm{O}(D)$-invariant tensor of rank $L=2k$.
Therefore, it is a linear combination of products of Kronecker deltas,
\begin{equation}\label{eq:theta}
	\kappa\big(X^{\mu_1}\cdots X^{\mu_{L}}\big)=\sum_{M}\theta_{M}\prod_{(s,t)\in M}\delta^{\mu_s\mu_t},
\end{equation}
where $M$ runs over the perfect matchings of the $L$ positions, as in~\eqref{eq:A-decomp}.
We label the coefficients $\theta_M$ in the same way as $A_M$, and call $L$ the length of $\theta_M$.
At length $4$, we have
\begin{equation}\label{eq:theta-4}
	\kappa\big(X^{\mu_1}X^{\mu_2}X^{\mu_3}X^{\mu_4}\big)=\theta_{1,2,3,4}\,\delta^{\mu_1\mu_2}\delta^{\mu_3\mu_4}+\theta_{1,4,2,3}\,\delta^{\mu_1\mu_4}\delta^{\mu_2\mu_3}+\theta_{1,3,2,4}\,\delta^{\mu_1\mu_3}\delta^{\mu_2\mu_4}.
\end{equation}
For concrete words, \eqref{eq:theta} keeps only the matchings that pair equal letters, so
\begin{equation}
	\kappa(XXYY)=\theta_{1,2,3,4},\qquad \kappa(YXXY)=\theta_{1,4,2,3},\qquad \kappa(XYXY)=\theta_{1,3,2,4}.
\end{equation}
As for the moments, the cyclicity of the trace gives $\kappa(YXXY)=\kappa(XXYY)$, so $\theta_{1,4,2,3}=\theta_{1,2,3,4}$.
In general, $\theta_M$ depends only on the class $[M]$ defined in section~\ref{sec:ODdecomposition},
and at sufficient large $D$ the independent parameters are $\theta_{[M]}$, one for each class.
The numbers of classes are
\begin{equation}\label{eq:counting}
	1,\ 2,\ 5,\ 17,\ 79,\ 554\qquad \text{for}\qquad L=2,4,6,8,10,12 .
\end{equation}
They are the same as the numbers of the independent coefficients $A$~\cite{Li:2025ymi},
since~\eqref{eq:theta} is the analogue of~\eqref{eq:A-decomp} for the free cumulants.

In terms of the parameters, the free cumulants~\eqref{eq:abc} are $a=\theta_{1,2}$, $b=\theta_{1,2,3,4}$ and $c=\theta_{1,3,2,4}$.
At length $6$, there are five classes, and each of them is selected by exactly one word.
We define the five length-6 parameters $u_1,\dots,u_5$ as the free cumulants of these words:
\begin{equation}\label{eq:u}
	\begin{gathered}
		u_1\equiv\kappa(XXYYZZ),\qquad u_2\equiv\kappa(XXYZYZ),\qquad u_3\equiv\kappa(XXYZZY),\\
		u_4\equiv\kappa(XYXZYZ),\qquad u_5\equiv\kappa(XYZXYZ),
	\end{gathered}
\end{equation}
where $X,Y,Z$ are distinct matrices.
In the labelling above, they are $u_1=\theta_{1,2,3,4,5,6}$, $u_2=\theta_{1,2,3,5,4,6}$, $u_3=\theta_{1,2,3,6,4,5}$, $u_4=\theta_{1,3,2,5,4,6}$ and $u_5=\theta_{1,4,2,5,3,6}$,
as listed in table~\ref{tab:low-cumulants}.
Table~\ref{tab:low-cumulants} expresses $a,b,c,u_1,\dots,u_5$ in terms of moments.
Other words of length $\leq 6$ are polynomials in these eight parameters.
For instance,
\begin{equation}\label{eq:mom-cum-6}
	\begin{aligned}
		\avg{XXXX}&=2a^2+2b+c,\\
		\avg{XXXXYY}&=2a^3+5ab+ac+u_1+u_2+u_3,\\
		\avg{XXXYXY}&=2ac+2u_2+u_4,\\
		\avg{XXYXXY}&=a^3+2ab+u_3+u_4+u_5 .
	\end{aligned}
\end{equation}

\begin{table}[htbp]
	\centering
	\renewcommand{\arraystretch}{1.25}
	\begin{tabular}{>{$}l<{$} >{$}l<{$} >{$}l<{$} >{$}l<{$}}
		\toprule
		\text{parameter} & \text{free cumulant} & \text{in terms of moments} & \theta_M\\
		\midrule
		a & \kappa(XX) & \avg{XX} & \theta_{1,2}\\
		b & \kappa(XXYY) & \avg{XXYY}-a^2 & \theta_{1,2,3,4}\\
		c & \kappa(XYXY) & \avg{XYXY} & \theta_{1,3,2,4}\\
		u_1 & \kappa(XXYYZZ) & \avg{XXYYZZ}-a^3-3ab & \theta_{1,2,3,4,5,6}\\
		u_2 & \kappa(XXYZYZ) & \avg{XXYZYZ}-ac & \theta_{1,2,3,5,4,6}\\
		u_3 & \kappa(XXYZZY) & \avg{XXYZZY}-a^3-2ab & \theta_{1,2,3,6,4,5}\\
		u_4 & \kappa(XYXZYZ) & \avg{XYXZYZ} & \theta_{1,3,2,5,4,6}\\
		u_5 & \kappa(XYZXYZ) & \avg{XYZXYZ} & \theta_{1,4,2,5,3,6}\\
		\bottomrule
	\end{tabular}
	\caption{The parameters of the free cumulants of length $2,4,6$.
		The third column follows from~\eqref{eq:mom-cum}.
		The last column gives the parameter $\theta_M$ in~\eqref{eq:theta}.
		It has the same subscripts as the corresponding moment coefficient $A$ in~\eqref{eq:A-decomp},
		but it is the connected part, e.g., $b=\theta_{1,2,3,4}=A_{1,2,3,4}-A_{1,2}^2$.}
	\label{tab:low-cumulants}
\end{table}

Finally, the free cumulants of a single matrix follow from~\eqref{eq:theta} by setting all the indices equal,
\begin{equation}\label{eq:single-kappa}
	\kappa_L(X)\equiv\kappa(\underbrace{X\cdots X}_{L})=\sum_{M} \theta_{M},
\end{equation}
where the sum runs over all the perfect matchings of $L$ positions.
For example, $\kappa_2(X)=a$ and $\kappa_4(X)=2b+c$.
They will be used in section~\ref{sec:trajectory}.

\subsection{Truncation and loop equations}
\label{sec:truncation}

Now we introduce the only approximation.
For an even cutoff $K$, we set
\begin{equation}\label{eq:truncation}
	\theta_M=0\qquad\text{for}\qquad L>K,
\end{equation}
i.e., all the free cumulants of length greater than $K$ vanish.
We refer to this approximation as the truncation $K$.
Then every moment is a polynomial in the remaining parameters $\theta_M$ of length $L\leq K$, given by~\eqref{eq:mom-cum}.

Next, we take all the loop equations~\eqref{eq:loop} with words of length $\leq K+2$,
i.e., with $\mathcal{O}$ of length $\leq K-1$.
We express them in terms of $\theta_M$, and remove the  duplicate equations.
This gives a system of polynomial equations for the parameters.
The words of length $K+2$ contain free cumulants of length $K+2$, which are set to zero by the truncation.
Therefore, the longest loop equations are the ones where the truncation enters.
With this choice, the number of independent equations equals the number of unknowns at $K=4$ and exceeds it for $K\geq 6$, as we will see below.

\paragraph{Why the truncation is reasonable.}\label{sec:why}
The truncation is motivated by the large $D$ expansion.
At the leading order, the matrices $X^\mu$ are independent Gaussian matrices~\cite{Hotta:1998en},
so all the free cumulants except $\kappa(XX)$ vanish, as discussed in section~\ref{sec:cumulant-def}.
From the derivation in \cite{Hotta:1998en},  we can see that the longer free cumulants are suppressed by higher powers of $1/D$,
\begin{equation}\label{eq:power-counting}
	\kappa\big(X^{\mu_1}\cdots X^{\mu_{2k}}\big)=O\Big(\avg{XX}^k\,D^{-(k-1)}\Big).
\end{equation}
We do not derive~\eqref{eq:power-counting}, but we can check it against the large $D$ results for concrete words in~\cite{Li:2025ymi}.
For example, the moment-cumulant formula~\eqref{eq:mom-cum} gives
\begin{equation}
	\avg{XXYYZZ}=\avg{XX}^3+3\avg{XX}\big(\avg{XXYY}-\avg{XX}^2\big)+\kappa(XXYYZZ).
\end{equation}
The large $D$ results~\cite{Li:2025ymi}
\begin{equation}
	\begin{aligned}
		\avg{XX}&=\tfrac{1}{\sqrt{2}}D^{-1/2}+\tfrac{7}{6\sqrt{2}}D^{-3/2}+O(D^{-5/2}),\\
		\avg{XXYY}&=\tfrac12 D^{-1}+D^{-2}+O(D^{-3}),\\
		\avg{XXYYZZ}&=\tfrac{1}{2\sqrt{2}}D^{-3/2}+\tfrac{5}{4\sqrt{2}}D^{-5/2}+O(D^{-7/2})
	\end{aligned}
\end{equation}
show that the first two terms on the right-hand side reproduce $\avg{XXYYZZ}$ up to the subleading order,
so $\kappa(XXYYZZ)=O(D^{-7/2})$, in agreement with~\eqref{eq:power-counting} for $k=3$.
More generally, as we will see in section~\ref{sec:largeD}, the $K=4$ solution reproduces the leading and subleading coefficients of all the words listed in~\cite{Li:2025ymi} (table~\ref{tab:K4-largeD}),
so the free cumulants of length $\geq 6$ do not contribute at these orders. The large $D$ coefficients in section~\ref{sec:largeD} show that this pattern continues to higher orders.
At small $D$, the suppression is weak, and we expect a slower convergence.

\paragraph{Truncation as an ansatz for analytic trajectories.}\label{sec:trajectory}
For a single matrix, the truncation has a simple interpretation in terms of the resolvent.
Let
\begin{equation}
	G(z)=\Big\langle \tr\,\frac{1}{z-X}\Big\rangle=\sum_{L\geq 0}\avg{X^L}\,z^{-L-1}
\end{equation}
be the resolvent of a single matrix, and let
\begin{equation}\label{eq:Rtransform}
	\mathcal{R}(w)=\sum_{L\geq 1}\kappa_L(X)\,w^{L-1}
\end{equation}
be its R-transform, where $\kappa_L(X)$ is given by~\eqref{eq:single-kappa}.
The moment-cumulant formula~\eqref{eq:mom-cum} for a single matrix is equivalent to~\cite{Nica:2006}
\begin{equation}\label{eq:G-R}
	z=\mathcal{R}\big(G(z)\big)+\frac{1}{G(z)} .
\end{equation}
At $K=2$, $\mathcal{R}(w)=a\,w$, and $G(z)$ corresponds to Wigner's semicircle.
For finite $K$, $\mathcal{R}$ is a polynomial of degree $K-1$,
so $G(z)$ is a root of a polynomial equation of degree $K$.
In other words, the truncation in the cumulants is a resummation of the moments:
the infinite set of moments $\avg{X^L}$ is determined by finitely many parameters.
In the language of the analytic trajectory bootstrap, the moments $\avg{X^L}$ form a trajectory that is analytic in $L$, given by the eigenvalue density computed in section~\ref{sec:density}.
In the ansatz of~\cite{Li:2025ymi,Li:2024ggr}, the eigenvalue density near the endpoints is a square root times a finite power series.
In the free cumulant ansatz, the density is instead determined by a polynomial equation of degree $K$.

\paragraph{Comparison with previous bootstrap procedures.}\label{sec:consistency}

In the analytic trajectory bootstrap~\cite{Li:2024ggr}, the ansatz describes trajectories, the families of words (for example $\avg{X^{n}}$ and $\avg{X^{n-2}Y^2}$), as analytic functions of the lengths.
Besides the loop equations, such an ansatz has to satisfy several consistency conditions:
\begin{enumerate}
	\item \emph{Cyclicity and reflection.} Words related by the cyclicity of the trace or by reversing the order of the letters have the same expectation value.
	\item \emph{Symmetry.} Here, the moments must be $\mathrm{O}(D)$ covariant.
	\item \emph{Contraction limit.} When one of the lengths of a multi-length moment goes to zero, the moment must reduce to a moment with fewer lengths.
	For example, $\avg{X^{n_1}Y^{n_2}}$ must reduce to $\avg{X^{n_1}}$ at $n_2=0$.
	\item \emph{Intersections of trajectories.} Different families can contain the same word.
	For example, $\avg{X^{n}}$ and $\avg{X^{n-2}Y^2}$ must agree at $n=2$,
	and $\avg{X^{n-2}Y^2}$ and $\avg{X^{n-4}YX^2Y}$ must agree at $n=4$.
\end{enumerate}
These conditions constrain the parameters of the ansatz~\cite{Li:2025ymi}.
In our work, the free cumulant ansatz satisfies all these conditions automatically.
All the moments are computed from the same parameters $\theta_M$ by the moment-cumulant formula~\eqref{eq:mom-cum}.
Therefore, the intersections and the contraction limits hold identically.
The symmetries hold by construction~\eqref{eq:theta}.

As discussed above, the truncation can also be regarded as an ansatz for the trajectories.
In~\cite{Li:2025ymi}, the ansatz for the trajectories was motivated by the $1/D$ expansion,
but we had no systematic way to improve it to the next order.
Here, the ansatz can be improved systematically by increasing $K$.

\subsection{Solving the equations}
\label{sec:high}

For $K=2$ and $K=4$, the number of independent loop equations equals the number of unknowns,
and the system can be solved in closed form.
We give these solutions in section~\ref{sec:low}.
For $K\geq 6$, the loop equations of length $K+2$ provide more equations than unknowns.
For example, at $K=6$ there are $8$ unknowns and $23$ distinct loop equations of length $\leq 8$.
Since the truncation is an approximation, these equations cannot be satisfied exactly.
We minimize the sum of squared residuals
\begin{equation}\label{eq:eta}
	\eta=\sum_j \big(\text{loop eq}\big)_j^2 ,
\end{equation}
as in~\cite{Li:2025ymi}, where $j$ labels the loop equations and $(\text{loop eq})_j$ is the difference between the two sides of the $j$-th equation.
The polynomial system has several solutions, and the least squares problem has several local minima.
We select the solution by continuation in $K$.
At $K=4$, we take the closed-form solution with $\avg{XX}>0$ (section~\ref{sec:low}), which approaches the free solution as $D\to\infty$.
For $K\geq 6$,
the initial point is that the solution at $K-2$ for the same $D$, and the new parameters of length $K$ are set to zero.
Without this continuation, the solver can converge to a wrong local minimum.
For example, at $D=3$ and $K=10$, starting from the free solution, it converges to $\avg{XX}\approx 0.42$, far from the MC estimate $0.80$. 
The numbers of unknowns and equations are given in appendix~\ref{app:cost}.

This step differs from previous works.
In~\cite{Schwartz:1999cm}, the number of independent equations was found to be equal to the number of unknowns at the fourth order,
and the sixth order system with $9$ equations and $9$ unknowns was solved only at large $D$.\footnote{Schwartz did not impose the $\mathrm{O}(D)$ symmetry. His unknowns are labelled by index patterns such as $(1111)$, $(1122)$, $(1212)$, so his counting differs from ours.
	The results agree: our $K=6$ solution reproduces his large $D$ formula, see section~\ref{sec:largeD}.}
In~\cite{Frenkel:2026mqm}, one equation is selected for each unknown, and the square system is solved by Newton's method.

\section{Results}
\label{sec:results}

We now present the results of the free cumulant ansatz.
The lowest truncations $K=2$ and $K=4$ can be solved in closed form.
For $K=6,\dots,12$, the results are obtained numerically with the procedure of section~\ref{sec:high}:
we minimize $\eta$ in~\eqref{eq:eta} while following the solution in $K$.
In section~\ref{sec:finiteD}, we present $\avg{XX}$ and $\avg{XXXX}$ at finite $D$, from $K=2$ to $K=12$,
and compare them with the positivity bounds and Monte Carlo estimates.
Then we discuss other words in section~\ref{sec:other} and the eigenvalue densities in section~\ref{sec:density}.
Finally, in section~\ref{sec:largeD}, we study the large $D$ expansion.
We obtain the coefficients of the $1/D$ expansion of $\sqrt{2D}\,\avg{XX}$ both numerically and analytically,
and find the exact coefficients up to $D^{-4}$, two orders beyond the result of Schwartz~\cite{Schwartz:1999cm}.

\subsection{\texorpdfstring{$\langle \tr XX\rangle$ and $\langle \tr XXXX\rangle$}{<tr XX> and <tr XXXX>}}
\label{sec:finiteD}

We first solve the lowest truncations $K=2$ and $K=4$, where everything can be done by hand.

\paragraph{$K=2$: free matrices.}\label{sec:low}
At $K=2$, all the free cumulants of length $\geq 4$ vanish.
This means that the matrices are free, as in~\eqref{eq:freeness}, and that each of them has the semicircle distribution of a Gaussian matrix.
The only unknown is $a=\avg{XX}$, and we use the loop equations of length $\leq 4$.
There is only one, namely~\eqref{eq:loop-X}.
The moment-cumulant formula gives $\avg{XXYY}=a^2$ and $\avg{XYXY}=0$, so~\eqref{eq:loop-X} becomes
\begin{equation}\label{eq:L-K2}
	(D-1)\,a^2=\tfrac12 .
\end{equation}
The solution is
\begin{equation}\label{eq:K2}
	\avg{XX}=\frac{1}{\sqrt{2(D-1)}},\qquad \avg{XXXX}=2a^2=\frac{1}{D-1}.
\end{equation}
This solution is exact at the leading order of the $1/D$ expansion. This is the leading ansatz solution of the analytic trajectory bootstrap in~\cite{Li:2025ymi},
and the $n=2$ result of Schwartz~\cite{Schwartz:1999cm}.
It also coincides with the special point that saturates the $L_{\max}=6$ positivity bound in~\cite{Li:2025ymi}.

\paragraph{$K=4$: closed-form solution.}
At $K=4$, the unknowns are $a$ and the length-4 parameters $b,c$ of table~\ref{tab:low-cumulants}, and the length-6 parameters $u_1,\dots,u_5$ vanish.
There are four distinct loop equations of length $\leq 6$, and three of them are independent~\cite{Li:2025ymi}.
We use these three equations.
Using table~\ref{tab:low-cumulants} and~\eqref{eq:mom-cum-6} with $u_i=0$, they read
\begin{align}
	\text{(L1)}\quad \mathcal{O}=X,\ X^\nu=X:&\qquad (D-1)(a^2+b-c)=\tfrac12,\label{eq:L1}\\
	\text{(L2)}\quad \mathcal{O}=XYX,\ X^\nu=Y:&\qquad a\,(a^2+2b-Dc)=0,\label{eq:L2}\\
	\text{(L3)}\quad \mathcal{O}=XXX,\ X^\nu=X:&\qquad (D-1)(2a^3+5ab-ac)-a=0.\label{eq:L3}
\end{align}
Equation~\eqref{eq:L1} is~\eqref{eq:loop-X}, while at $K=2$ it reduces to~\eqref{eq:L-K2}.
These are three equations for the three unknowns $a,b,c$.
The result is
\begin{equation}\label{eq:K4}
	a^2=\frac{3D+2}{2(D-1)(3D-2)},\qquad b=-\frac{1}{2(D-1)(3D-2)},\qquad c=\frac{3}{2(D-1)(3D-2)},
\end{equation}
so that
\begin{equation}\label{eq:K4-X4}
	\avg{XX}=a=\sqrt{\frac{3D+2}{2(D-1)(3D-2)}},\qquad
	\avg{XXXX}=2a^2+2b+c=\frac{6D+5}{2(D-1)(3D-2)} .
\end{equation}
The expression for $\avg{XX}$ is the $n=4$ result of Schwartz,
$[2(D-1)(1-4/(3D+2))]^{-1/2}$~\cite{Schwartz:1999cm}.

At finite $D$, the $K=4$ solution is less accurate.
For $D=10$, we have $a=0.2520$, which is $1.3\%$ below the MC estimate $0.2552$.
For $D=4$, the deviation is $10\%$.

\paragraph{Higher truncations.}
For $K\geq 6$, the loop equations cannot be solved exactly, and we minimize $\eta$ in~\eqref{eq:eta} as described in section~\ref{sec:high}.
For each $D$, the solution at $K-2$ is used as the initial point for $K$, starting from the closed-form solution at $K=4$.

\begin{table}[htbp]
	\centering
	\small
	\setlength{\tabcolsep}{4.2pt}
	\begin{tabular}{cccccccccc}
		\toprule
		\multicolumn{10}{c}{$\avg{XX}$}\\
		\midrule
		$D$ & $K=2$ & $K=4$ & $K=6$ & $K=8$ & $K=10$ & $K=12$ & dev. & positivity & MC\\
		\midrule
		3 & 0.5000 & 0.6268 & 0.6716 & 0.6978 & 0.7164 & 0.7302 & $-9.0\%$ & 0.85(14) & 0.8028\\
		4 & 0.4082 & 0.4830 & 0.5077 & 0.5199 & 0.5265 & 0.5306 & $-1.5\%$ & 0.540(26) & 0.5386\\
		5 & 0.3536 & 0.4043 & 0.4189 & 0.4248 & 0.4274 & 0.4287 & $-0.39\%$ & 0.43(1) & 0.4304\\
		6 & 0.3162 & 0.3536 & 0.3629 & 0.3661 & 0.3672 & 0.3677 & $-0.16\%$ & 0.368(5) & 0.3683\\
		7 & 0.2887 & 0.3176 & 0.3240 & 0.3258 & 0.3264 & 0.3266 & $-0.05\%$ & 0.3265(27) & 0.3268\\
		8 & 0.2673 & 0.2905 & 0.2951 & 0.2963 & 0.2966 & 0.2967 & $-0.10\%$ & 0.2965(18) & 0.2970\\
		9 & 0.2500 & 0.2693 & 0.2727 & 0.2734 & 0.2736 & 0.2737 & $-0.01\%$ & 0.2735(12) & 0.2737\\
		10 & 0.2357 & 0.2520 & 0.2546 & 0.2551 & 0.2552 & 0.2553 & $+0.02\%$ & 0.25515(95) & 0.2552\\
		\midrule
		\multicolumn{10}{c}{$\avg{XXXX}$}\\
		\midrule
		$D$ & $K=2$ & $K=4$ & $K=6$ & $K=8$ & $K=10$ & $K=12$ & dev. & positivity & MC\\
		\midrule
		3 & 0.5000 & 0.8214 & 0.9645 & 1.0492 & 1.1052 & 1.1475 & $-16\%$ & 1.61(54) & 1.367\\
		4 & 0.3333 & 0.4833 & 0.5391 & 0.5661 & 0.5808 & 0.5895 & $-2.8\%$ & 0.613(60) & 0.6065\\
		5 & 0.2500 & 0.3365 & 0.3634 & 0.3739 & 0.3786 & 0.3809 & $-0.70\%$ & 0.385(18) & 0.3836\\
		6 & 0.2000 & 0.2563 & 0.2711 & 0.2760 & 0.2778 & 0.2785 & $-0.36\%$ & 0.2798(78) & 0.2795\\
		7 & 0.1667 & 0.2061 & 0.2152 & 0.2177 & 0.2185 & 0.2188 & $-0.19\%$ & 0.2191(42) & 0.2192\\
		8 & 0.1429 & 0.1721 & 0.1780 & 0.1794 & 0.1798 & 0.1799 & $-0.26\%$ & 0.1801(23) & 0.1804\\
		9 & 0.1250 & 0.1475 & 0.1516 & 0.1524 & 0.1527 & 0.1527 & $-0.13\%$ & 0.1528(17) & 0.1529\\
		10 & 0.1111 & 0.1290 & 0.1319 & 0.1324 & 0.1326 & 0.1326 & $-0.15\%$ & 0.1326(11) & 0.1328\\
		\bottomrule
	\end{tabular}
	\caption{Results of the free cumulant ansatz for $\avg{XX}$ and $\avg{XXXX}$ at cutoffs $K=2,\dots,12$.
		The column ``dev.'' is the relative deviation of the $K=12$ result from the Monte Carlo (MC) estimate.
		The positivity intervals at $L_{\max}=12$ and the MC estimates are taken from~\cite{Li:2025ymi}.}
	\label{tab:XX}
\end{table}

\begin{figure}[htbp]
	\centering
	\includegraphics[width=0.78\textwidth]{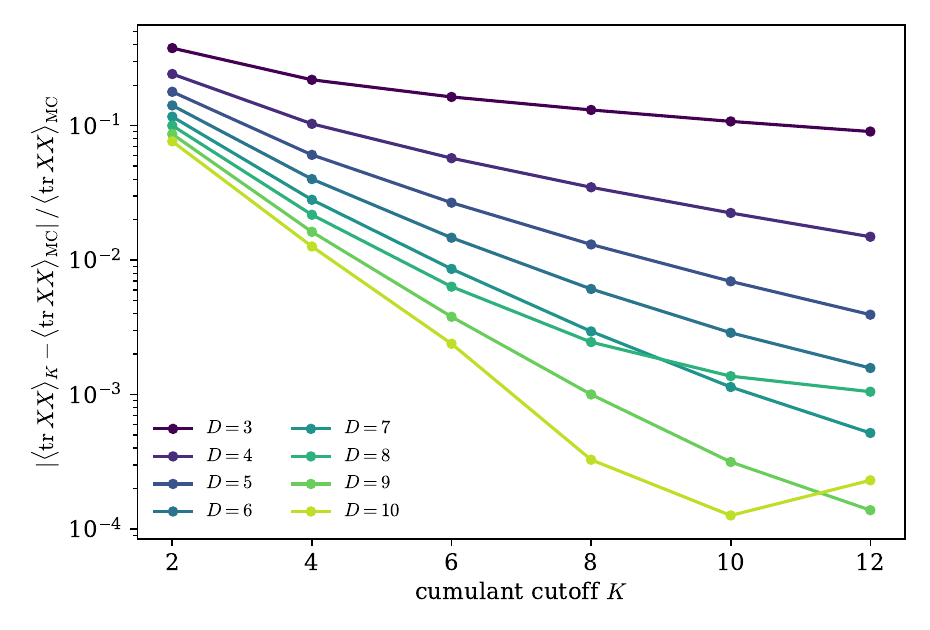}
	\caption{Relative deviation of $\avg{XX}$ from the Monte Carlo estimate as a function of the cumulant cutoff $K$, for $D=3,\dots,10$.
		The relative uncertainties of the Monte Carlo estimates are about $0.1\%$--$0.7\%$~\cite{Li:2025ymi},
		so deviations below $\sim 10^{-3}$ are not resolved.}
	\label{fig:convergence}
\end{figure}

Table~\ref{tab:XX} shows the results for $\avg{XX}$ and $\avg{XXXX}$ at $D=3,\dots,10$,
and figure~\ref{fig:convergence} shows the convergence of $\avg{XX}$ with $K$.
We make the following observations:
\begin{itemize}
	\item For every $D$, the results increase monotonically with $K$ and approach the Monte Carlo estimates from below (except $K=12, D=10$).\footnote{	We do not have a proof that this behavior persists at higher $K$,
		so the truncation results should not be interpreted as lower bounds.}
	
	\item The convergence is faster at larger $D$.
	Let $q_K$ be the result for a given observable at cutoff $K$, and let $\delta_K=q_K-q_{K-2}$ be the change when $K$ is increased by $2$.
	For $\avg{XX}$, the ratio $\delta_{10}/\delta_{12}$ is about $2.0$ for $D=5$, $1.6$ for $D=4$ and $1.3$ for $D=3$, and it is larger for larger $D$.
	\item At $K=12$, all the results lie inside the projections of the $L_{\max}=12$ positivity islands onto the two axes. 
	The deviations from the Monte Carlo estimates are smaller than the half widths of the positivity intervals.
	For $D\geq 7$, the deviations of $\avg{XX}$ are also smaller than the Monte Carlo uncertainties.
	\item For $D=3$, the convergence is slow.
	At $K=12$, $\avg{XX}$ and $\avg{XXXX}$ are still $9\%$ and $16\%$ below the Monte Carlo estimates.
\end{itemize}

\subsection{Other words}
\label{sec:other}

The solution at a given $K$ determines all the moments through~\eqref{eq:mom-cum}.
For $D=10$ and $K=12$, we obtain
\begin{equation}\label{eq:higher-moments}
	\avg{X^6}=0.0867,\qquad \avg{X^8}=0.0637,\qquad \avg{X^{10}}=0.0503 ,
\end{equation}
which are compatible with the Monte Carlo estimates $0.089(6)$, $0.066(9)$ and $0.052(5)$ in~\cite{Li:2025ymi}.

Table~\ref{tab:words} compares some longer words with Monte Carlo data at $K=12$.  The MC estimates in this table are from our new runs with an adapted version of the code of~\cite{Jha:2021exo}, at $N=200, 300, \dots,600$ for $D=4$ and $N=200,300$ for $D=10$, without extrapolation in $N$. Note that the Monte Carlo estimates and their error bars are very rough, because the Monte Carlo data fluctuate strongly.
For $D=10$, the deviations from the Monte Carlo estimates are at most $1.3$ times the estimated errors.
For $D=4$, the $K=12$ results are systematically below the Monte Carlo estimates by about $3\%$--$10\%$.

\begin{table}[htbp]
	\centering
	\small
	\begin{tabular}{lcccc}
		\toprule
		& \multicolumn{2}{c}{$D=4$} & \multicolumn{2}{c}{$D=10$}\\
		\cmidrule(lr){2-3}\cmidrule(lr){4-5}
		word & $K=12$ & MC & $K=12$ & MC\\
		\midrule
		$XXYY$ & 0.25205 & 0.2591(32) & 0.06272 & 0.0638(14)\\
		$XYXY$ & 0.08540 & 0.0925(31) & 0.00716 & 0.0083(14)\\
		$XXXXXX$ & 0.83505 & 0.885(44) & 0.08671 & 0.0933(90)\\
		$XXYYZZ$ & 0.10748 & 0.1128(97) & 0.01483 & 0.01544(60)\\
		$XXYZZY$ & 0.11840 & 0.1245(97) & 0.01542 & 0.01603(60)\\
		$XXYXXY$ & 0.13673 & 0.1451(49) & 0.01565 & 0.01631(86)\\
		$XXXYXY$ & 0.08615 & 0.0944(56) & 0.00358 & 0.0044(11)\\
		$XXYZYZ$ & 0.03713 & 0.0414(95) & 0.00170 & 0.00213(46)\\
		$XXXXXXXX$ & 1.33847 & 1.45(12) & 0.06373 & 0.072(11)\\
		$XXXXYXXY$ & 0.15154 & 0.165(10) & 0.00803 & 0.00865(84)\\
		$XXYYZZWW$ & 0.04081 & 0.0433(42) & 0.00337 & 0.00354(13)\\
		\bottomrule
	\end{tabular}
	\caption{Expectation values $\avg{\cdot}$ of some words at $K=12$, compared with Monte Carlo (MC) estimates. 
	}
	\label{tab:words}
\end{table}

\subsection{Eigenvalue density}
\label{sec:density}

We now compute the eigenvalue density of a single matrix from the resolvent~\eqref{eq:G-R} of section~\ref{sec:trajectory}.

\begin{figure}[htbp]
	\centering
	\includegraphics[width=0.78\textwidth]{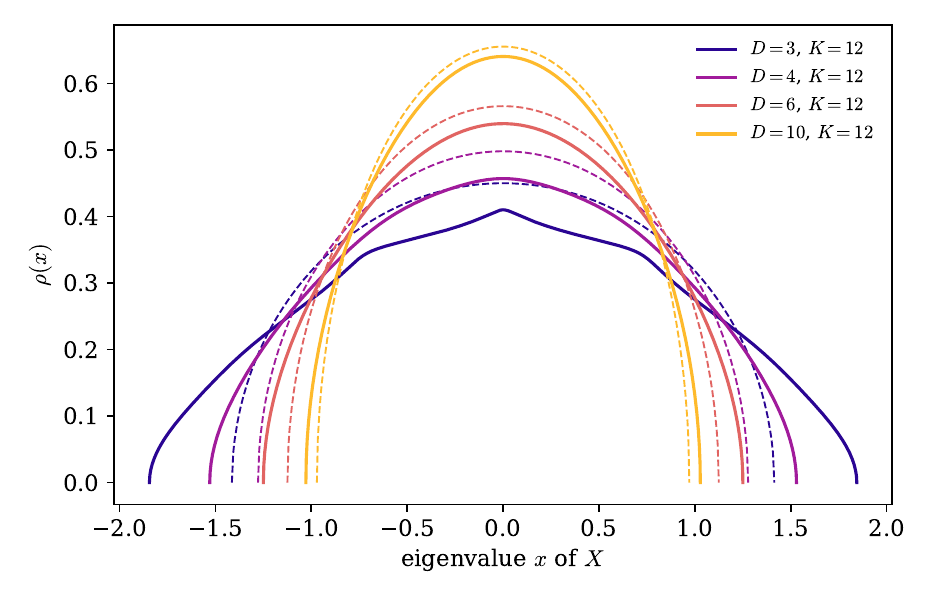}
	\caption{Eigenvalue densities $\rho(x)$ of a single matrix obtained from~\eqref{eq:G-R} at $K=12$ (solid curves),
		for $D=3,4,6,10$.
		The dashed curves are the semicircles of the $K=2$ truncation with the same $D$.}
	\label{fig:density}
\end{figure}

The physical branch of~\eqref{eq:G-R} satisfies $G(z)\sim 1/z$ at large $z$.
The endpoint of the eigenvalue distribution is $z_{\max}=\mathcal{R}(w_*)+1/w_*$,
where $w_*$ is the smallest positive root of $w^2\mathcal{R}'(w)=1$.
We follow the physical branch along the real axis and compute the density $\rho(x)=-\frac{1}{\pi}\,\mathrm{Im}\,G(x+i0)$.
Note that the truncation does not impose that $\rho(x)$ is a positive density.
For all $D=3,\dots,10$ and $K=2,\dots,12$, we find that $\rho(x)$ is positive and supported on a single interval $[-z_{\max},z_{\max}]$.
For $K=8,10,12$, we also checked that the integrals of $\rho(x)$, $x^2\rho(x)$ and $x^4\rho(x)$ reproduce $1$, $\avg{XX}$ and $\avg{XXXX}$ within the accuracy of the numerical integration ($< 10^{-5}$).
Figure~\ref{fig:density} shows the densities at $K=12$ for $D=3,4,6,10$, together with the semicircles at $K=2$.
At $D=10$, we obtain $z_{\max}=1.02802$, $1.02829$, $1.02835$ for $K=8,10,12$.
This is consistent with the analytic trajectory bootstrap result $z_{\max}\approx 1.0282$ in~\cite{Li:2025ymi}.

\subsection{\texorpdfstring{Large $D$}{Large D} expansion}
\label{sec:largeD}
In this subsection, we study the solutions at large $D$.
First, we expand the closed-form solution at $K=4$ (section~\ref{sec:low}), and find that it reproduces the leading and subleading orders of the $1/D$ expansion for all the words in~\cite{Li:2025ymi}.
Second, for higher truncations, we solve the truncated equations numerically at very large $D$ and extract the $1/D$ coefficients of $\avg{XX}$ by polynomial interpolation.
Each increase of $K$ gives one more coefficient.
Finally, we derive these coefficients analytically.
For this purpose, we keep all the free cumulants of length $\leq K'$ and use only the loop equations whose words have length $\leq K'$; these equations do not contain longer free cumulants.
Instead of the truncation, we impose an ansatz for the large $D$ form of these free cumulants, and solve the loop equations order by order in $1/D$.

\paragraph{The $K=4$ solution.}
We first expand the $K=4$ solution~\eqref{eq:K4} at large $D$ as
\begin{equation}\label{eq:K4-largeD}
	\avg{X^{\mu_1}X^{\mu_2}\cdots X^{\mu_L}}=D^{-L/4}\left(c_0+\frac{c_1}{D}+\frac{c_2}{D^2}+O(D^{-3})\right).
\end{equation}
The moments are computed from~\eqref{eq:K4} with the moment-cumulant formula~\eqref{eq:mom-cum}.
In table~\ref{tab:K4-largeD}, we list the coefficients for the $17$ words in section 2.3 of~\cite{Li:2025ymi}.
For the first $12$ words, the coefficients $c_0$ and $c_1$ agree with~\cite{Li:2025ymi}.
For the last $5$ words, we have $c_0=0$, only $c_1$ is given in~\cite{Li:2025ymi}, and it also agrees.
Therefore, the $K=4$ truncation is exact up to the subleading order in $1/D$.
The coefficients $c_2$ are not exact in general.
For example, $c_2=67/(72\sqrt2)$ for $\avg{XX}$, while the exact value is $\gamma_2/\sqrt2$ with $\gamma_2=1343/648$, see~\eqref{eq:gamma2} below.
At the same time, the $K=4$ solution is a closed-form function of $D$, which can be evaluated at any finite $D$, see section~\ref{sec:finiteD}.

\begin{table}[htbp]
	\centering
	\renewcommand{\arraystretch}{1.35}
	\begin{tabular}{lccc}
		\hline
		word & $c_0$ & $c_1$ & $c_2$ ($K=4$)\\
		\hline
		$\avg{X^2}$ & $\frac{1}{\sqrt2}$ & $\frac{7}{6\sqrt2}$ & $\frac{67}{72\sqrt2}$\\
		$\avg{X^4}$ & $1$ & $\frac{5}{2}$ & $\frac{7}{2}$\\
		$\avg{X^6}$ & $\frac{5}{2\sqrt2}$ & $\frac{39}{4\sqrt2}$ & $\frac{961}{48\sqrt2}$\\
		$\avg{X^8}$ & $\frac{7}{2}$ & $\frac{56}{3}$ & $\frac{919}{18}$\\
		$\avg{X^2Y^2}$ & $\frac{1}{2}$ & $1$ & $\frac{4}{3}$\\
		$\avg{X^4Y^2}$ & $\frac{1}{\sqrt2}$ & $\frac{19}{6\sqrt2}$ & $\frac{427}{72\sqrt2}$\\
		$\avg{X^6Y^2}$ & $\frac{5}{4}$ & $\frac{67}{12}$ & $\frac{124}{9}$\\
		$\avg{X^4Y^4}$ & $1$ & $\frac{17}{4}$ & $\frac{185}{18}$\\
		$\avg{X^2Y^2Z^2}$ & $\frac{1}{2\sqrt2}$ & $\frac{5}{4\sqrt2}$ & $\frac{97}{48\sqrt2}$\\
		$\avg{X^4Y^2Z^2}$ & $\frac{1}{2}$ & $\frac{7}{4}$ & $\frac{65}{18}$\\
		$\avg{X^2YZ^2Y}$ & $\frac{1}{2\sqrt2}$ & $\frac{17}{12\sqrt2}$ & $\frac{359}{144\sqrt2}$\\
		$\avg{X^4YZ^2Y}$ & $\frac{1}{2}$ & $2$ & $\frac{55}{12}$\\\hline
		$\avg{(XY)^2}$ & $0$ & $\frac{1}{2}$ & $\frac{5}{6}$\\
		$\avg{X^2(XY)^2}$ & $0$ & $\frac{1}{\sqrt2}$ & $\frac{17}{6\sqrt2}$\\
		$\avg{X^4(XY)^2}$ & $0$ & $\frac{5}{4}$ & $\frac{31}{6}$\\
		$\avg{X^2(YZ)^2}$ & $0$ & $\frac{1}{2\sqrt2}$ & $\frac{17}{12\sqrt2}$\\
		$\avg{X^4(YZ)^2}$ & $0$ & $\frac{1}{2}$ & $\frac{25}{12}$\\
		\hline
	\end{tabular}
	\caption{Large $D$ coefficients~\eqref{eq:K4-largeD} of the $K=4$ solution.
		The coefficients $c_0$ and $c_1$ agree with the large $D$ results in~\cite{Li:2025ymi}.
		For the last $5$ words, only $c_1$ is known in~\cite{Li:2025ymi}.
		The coefficients $c_2$ are the predictions of the $K=4$ truncation, and they are not exact.}
	\label{tab:K4-largeD}
\end{table}

\paragraph{Higher truncations: numerical interpolation.}

To study the large $D$ behavior, we solve the truncated system numerically at the six values $D=10^4\times 2^j$, $j=0,\dots,5$, with $50$-digit precision.
We use the truncations up to $K=10$.
We do not include $K=12$, because solving the $K=12$ system with this precision at many values of $D$ is slow.
As in section~\ref{sec:high}, the solution at $K$ is obtained from the solution at $K-2$, starting from the closed-form solution at $K=4$.
Then we extract the coefficients of
\begin{equation}\label{eq:gamma}
	\sqrt{2D}\,\avg{XX}=1+\frac{\gamma_1}{D}+\frac{\gamma_2}{D^2}+\frac{\gamma_3}{D^3}+\cdots
\end{equation}
by polynomial interpolation of degree $5$ in $1/D$ through the six points.
The results are shown in table~\ref{tab:gamma}.
For $K=4$, this procedure reproduces the exact coefficients $\gamma_1,\gamma_2,\gamma_3$ to $10^{-12}$ and $\gamma_4$ to $10^{-8}$.

\begin{table}[htbp]
	\centering
	\small
	\begin{tabular}{llcccc}
		\toprule
		& method & $\gamma_1$ & $\gamma_2$ & $\gamma_3$ & $\gamma_4$\\
		\midrule
		$K=2$ & closed form & $1/2$ & $3/8$ & $5/16$ & $35/128$\\
		$K=4$ & closed form & $\mathbf{7/6}$ & $67/72$ & $355/432$ & $2465/3456$\\
		$K=6$ & numerical & $\mathbf{1.1666666667}$ & $\mathbf{2.0725308642}$ & $1.2434$ & $-0.6553$\\
		$K=8$ & numerical & $\mathbf{1.1666666667}$ & $\mathbf{2.0725308642}$ & $\mathbf{3.4341278006}$ & $1.5664$\\
		$K=10$ & numerical & $\mathbf{1.1666666667}$ & $\mathbf{2.0725308642}$ & $\mathbf{3.4341278006}$ & $\mathbf{5.9998}$\\
		\midrule
		exact & analytic $1/D$ & $\frac{7}{6}$ & $\frac{1343}{648}$ & $\frac{120167}{34992}$ & $\frac{15116017}{2519424}$\\[2pt]
		& (decimal) & $1.1666666667$ & $2.0725308642$ & $3.4341278006$ & $5.9997908252$\\
		\bottomrule
	\end{tabular}
	\caption{Coefficients of the large $D$ expansion~\eqref{eq:gamma} of $\sqrt{2D}\,\avg{XX}$.
		Upper part: truncation $K$.
		For $K=2,4$, the coefficients are obtained by expanding the closed-form solutions, and they are exact rational numbers for these truncations.
		For $K=6,8,10$, they are obtained numerically by the polynomial interpolation described in the text.
		Lower part: the exact coefficients obtained from the analytic $1/D$ expansion~\eqref{eq:gamma-exact}, as fractions and as decimals.
		The coefficients in bold agree with the exact values to all the digits shown.}
	\label{tab:gamma}
\end{table}

Table~\ref{tab:gamma} shows a clear pattern:
the truncation $K=2k$ gives the exact coefficients $\gamma_1,\dots,\gamma_{k-1}$,
and the next coefficient $\gamma_k$ changes at $K=2k+2$.
This is consistent with the power counting~\eqref{eq:power-counting}:
the neglected cumulants of length $2k+2$ first contribute at order $D^{-k}$ relative to the leading term.
For instance, the values of $\gamma_3$ at $K=8$ and $K=10$ agree to $10^{-12}$.

The coefficient $\gamma_1=7/6$ agrees with~\eqref{eq:largeD-sub}.
For $\gamma_2$, the values at $K=6,8,10$ agree with the rational number
\begin{equation}\label{eq:gamma2}
	\gamma_2=\frac{1343}{648}=2.0725308641975\ldots
\end{equation}
to $10^{-12}$.
It agrees with the $1/D^2$ term of the sixth-order result of Schwartz~\cite{Schwartz:1999cm},
\begin{equation}\label{eq:schwartz6}
	\avg{XX}^{-2}=2(D-1)\left[1-\frac{4}{3D+2}-\frac{185}{81D^2}+O(D^{-3})\right].
\end{equation}
The coefficients $\gamma_3=3.4341278006\ldots$ and $\gamma_4\approx 5.9998$ are new.
Below we derive $\gamma_2$, $\gamma_3$ and $\gamma_4$ exactly.

\paragraph{Analytic $1/D$ expansion.}
The pattern in table~\ref{tab:gamma} can be turned into an exact computation.
As observed above, the truncation $K=2k$ reproduces the first $k$ orders of the $1/D$ expansion.
We use the lowest truncations to find the form of the free cumulants at large $D$.
Recall that, by~\eqref{eq:cum-recursion},
\begin{equation}
	\begin{aligned}
		a &\equiv\kappa(XX)=\avg{XX},\qquad
		b\equiv\kappa(XXYY)=\avg{XXYY}-\avg{XX}^2,\\
		c &\equiv\kappa(XYXY)=\avg{XYXY}.
	\end{aligned}
\end{equation}

At $K=2$, the only nonzero parameter is $a$.
Keeping only the leading order of~\eqref{eq:K2}, we have
\begin{equation}
	a=\frac{1}{\sqrt{2D}}+O(D^{-3/2}) .
\end{equation}

At $K=4$, the nonzero parameters are $a$, $b$ and $c$.
Keeping the first two orders of~\eqref{eq:K4}, we have
\begin{equation}
	a=\frac{1}{\sqrt{2D}}\Big(1+\frac{7}{6D}\Big)+O(D^{-5/2}),\qquad
	b=-\frac{1}{6D^{2}}+O(D^{-3}),\qquad
	c=\frac{1}{2D^{2}}+O(D^{-3}) .
\end{equation}

In this pattern, each new order brings in the free cumulants that are two letters longer,
and a free cumulant of length $L$ starts at the order $D^{1-3L/4}$, i.e., $D^{-1/2}$ for $a$ and $D^{-2}$ for $b$ and $c$.
The higher terms are suppressed by integer powers of $1/D$.
We therefore assume that, for every parameter of length $L$,
\begin{equation}\label{eq:pert-ansatz}
	\theta_M=D^{1-3L/4}\sum_{j\geq 0}\theta_M^{(j)}D^{-j},
\end{equation}
where the coefficients $\theta_M^{(j)}$ do not depend on $D$.
This agrees with the power counting from known large $D$ expansion results~\eqref{eq:power-counting}.

The ansatz~\eqref{eq:pert-ansatz} cannot be combined directly with the truncation we used above.
Instead, we keep all the parameters of length $\leq K'$ and use only the loop equations whose words have length $\leq K'$.
These equations contain no free cumulant of length $>K'$, so they hold exactly.
We substitute~\eqref{eq:pert-ansatz}, expand in $1/D$, and solve order by order.
With $K'=6,8,10,12$, we obtain $\gamma_1,\dots,\gamma_{K'/2-2}$:
\begin{equation}\label{eq:gamma-exact}
	\sqrt{2D}\,\avg{XX}=1+\frac{7}{6D}+\frac{1343}{648\,D^2}+\frac{120167}{34992\,D^3}+\frac{15116017}{2519424\,D^4}+O(D^{-5}).
\end{equation}
The last coefficient is $\gamma_4=5.9997908\ldots$.
This confirms the numerical values in table~\ref{tab:gamma}, including $\gamma_4$ at $K=10$, and the result of Schwartz~\eqref{eq:schwartz6}.
At $K'=12$, there are $658$ parameters, and the computation takes about 2 hours on a laptop.
For the other parameters, we find for example
\begin{equation}
	b=-\frac{1}{D^2}\Big(\frac16+\frac{95}{162\,D}+\frac{6427}{4374\,D^2}+\cdots\Big),\qquad
	c=\frac{1}{D^2}\Big(\frac12+\frac{5}{3D}+\frac{629}{162\,D^2}+\cdots\Big),
\end{equation}
and at the leading order $u_4=\frac{1}{2\sqrt2}D^{-7/2}$, while $u_5=O(D^{-9/2})$.

\section{Discussion}
\label{sec:discussion}

In this work, we studied the large $N$ bosonic Yang-Mills matrix integrals 
with the analytic trajectory bootstrap, using the free cumulant truncation of Schwartz as a new ansatz.
The unknowns are the free cumulants, and for finite $D$ the only approximation is to set the cumulants of length greater than $K$ to zero.
The loop equations then determine the remaining cumulants without using positivity.
The truncation $K=2$ is the free approximation, and $K=4$ gives a closed-form solution that is exact up to the subleading order in $1/D$.
At $K=12$, the results for $\avg{XX}$ for $D\geq 5$ are within $0.4\%$ of the Monte Carlo estimates,
and all the results lie inside the positivity intervals.
The truncated cumulants give a positive eigenvalue density.
At large $D$, the truncation $K=2k$ fixes the first $k$ orders of the $1/D$ expansion numerically.
Guided by this pattern, we imposed an ansatz for the large $D$ form of the free cumulants and solved the loop equations order by order in $1/D$,
and obtained the exact coefficients of $\sqrt{2D}\,\avg{XX}$ up to $D^{-4}$, see~\eqref{eq:gamma-exact}.

Below are several comments on the limitations of this work.
\begin{itemize}
	\item The results are point estimates without rigorous error bars.
	The change $\delta_K$ between successive cutoffs may underestimate the error at small $D$.
	It would be useful to derive an error estimate for this bootstrap method.
	\item For $D=3$, the convergence in $K$ is slow.
The eigenvalue density at $K=12$ has a sharp peak at $x=0$ and two shoulders, which are artifacts of the truncation, see figure~\ref{fig:density}.
	This is consistent with the observation in~\cite{Li:2025ymi} that small $D$ is more difficult for both bootstrap methods.
	The power counting~\eqref{eq:power-counting} is based on $1/D$, so it is not surprising that the truncation is less effective at $D=3$.

\end{itemize}

Some future directions are in order.
\begin{itemize}
	\item One of the main motivations of this work is the supersymmetric case.
	The method uses only the loop equations and does not impose positivity,
	so a Pfaffian that is not real and positive is not an obstacle in principle.

	\item The truncation and the positivity bootstrap can be combined.
	For example, one may impose the positivity of the moment matrix on the truncated solution,
	or use the truncated solution as a starting point of the semidefinite programming.  The positivity of the eigenvalue density is not imposed but satisfied in all cases studied here.
	
	\item The Yang-Mills matrix integral is, in some sense, the reduction of $D$-dimensional Yang-Mills theory to a single point.
	It is therefore natural to ask whether the method extends to Yang-Mills theory itself.
	For large $N$ lattice Yang-Mills theory, see the work of positivity bootstrap in~\cite{Kazakov:2022xuh,Kazakov:2024ool}.
\end{itemize}

\acknowledgments
I am very grateful to Wenliang Li for many valuable ideas, discussions and comments on the manuscript. The work has also benefited from the code and data obtained in our previous collaboration~\cite{Li:2025ymi}.
I used Claude Opus 5.5 (Anthropic) to generate portions of the
Mathematica code and to assist in drafting the manuscript.
In particular, Claude helped me locate the earlier work by
Schwartz~\cite{Schwartz:1999cm} and identify the connection between his
method and ours. I reviewed the code, checked the reported results,
and take responsibility for the final manuscript.

\appendix

\section{Sizes of the systems}
\label{app:cost}

Table~\ref{tab:cost} lists the numbers of parameters and loop equations at each cutoff,
together with rough computation times in \texttt{Mathematica} 13.1 on a laptop.
The unknowns are the parameters $\theta_M$ of length $L\leq K$.
The equations are the loop equations~\eqref{eq:loop} of length $\leq K+2$ for concrete words,
written in terms of the unknowns through~\eqref{eq:mom-cum}.
We count them after removing the equations that are related by the symmetries of section~\ref{sec:setup},
the equations that become trivial ($0=0$), and the duplicates.
The number of equations depends on $D$ for $D<K/2$,
because the word $\mathcal{O}X^\nu$ in a loop equation of length $K+2$ has length $K$ and can contain up to $K/2$ distinct letters.
The construction of the equations, i.e., the expansion of all words through~\eqref{eq:mom-cum}, dominates the cost.

\begin{table}[h]
	\centering
	\small
	\begin{tabular}{cccccc}
		\toprule
		$K$ & parameters & loop equations ($D\geq 6$) & construction & solution & rms residual ($D=10$)\\
		\midrule
		$4$ & $3$ & $4$ & seconds & seconds & exact\\
		$6$ & $8$ & $23$ & seconds & seconds & $7\times 10^{-6}$ \\
		$8$ & $25$ & $226$ & seconds & seconds & $4\times 10^{-7}$\\
		$10$ & $104$ & $3564$ & about a minute & seconds & $2\times 10^{-8}$\\
		$12$ & $658$ & $78\,994$ & hours & minutes & $8\times 10^{-10}$\\
		\bottomrule
	\end{tabular}
	\caption{Numbers of parameters and loop equations, rough computation times, and the root mean square of the residuals at $D=10$.
		The numbers of equations are the numbers of distinct equations, not of independent ones; for instance, three of the four equations at $K=4$ are independent. 		For $D<K/2$, some combinations of the parameters are redundant (appendix~\ref{app:redundancy}), and
		there are fewer equations:
		$170$ for $K=8$ and $D=3$;
		$1431$ and $3083$ for $K=10$ and $D=3,4$;
		$12\,594$, $47\,286$ and $73\,807$ for $K=12$ and $D=3,4,5$.
		At $K=12$, the residuals increase to $1.2\times 10^{-7}$, $1.3\times 10^{-5}$ and $8\times 10^{-4}$ for $D=6,4,3$.}
	\label{tab:cost}
\end{table}

\section{\texorpdfstring{Redundancy of the cumulant parameters at small $D$}{Redundancy of the cumulant parameters at small D}}
\label{app:redundancy}

The parametrization~\eqref{eq:theta} uses formal products of Kronecker deltas.
For small $D$, these products are not linearly independent.
A product of $m$ deltas has $2m$ indices,
and distinct pairings can be linearly dependent only if $m>D$~\cite{Li:2025ymi}.
As a result, at small $D$, some combinations of the parameters $\theta_{[M]}$ do not affect any concrete word.
This is the analogue of the redundancy of the coefficients $A$ discussed in appendix B of~\cite{Li:2025ymi}.

In our computations, the redundancy appears for the cumulants of length $8$ at $D=3$,
of length $10$ at $D=3,4$, and of length $12$ at $D=3,4,5$.
The least squares solver then returns one representative of a family of equivalent solutions.
In this representative, some parameters can be very large.
For example, at $K=12$ and $D=3$, some parameters are of order $10^5$.
The concrete words are not affected.
For instance, the single-matrix cumulants~\eqref{eq:single-kappa} are of order one,
and the moments reconstructed from them agree with the values computed directly to machine precision.
For $D\geq 6$, all the parameters are of the expected size,
e.g., the length-12 parameters are of order $10^{-8}$ at $D=10$.

\end{document}